\documentclass{JAC2000}

\usepackage{graphicx}
\usepackage{epsfig}
\usepackage{graphicx}

\begin{document}

\title{Measurement of the six Dimensional Phase Space\\ at the New GSI High
Current Linac}
\author{P. Forck, F. Heymach, T. Hoffmann, A. Peters, P. Strehl\\
Gesellschaft f\"ur Schwerionenforschung GSI, Planck Strasse 1, 64291
Darmstadt, Germany\\
e-mail: p.forck@gsi.de}

\maketitle

\begin{abstract} 

{\normalsize
For the characterization of the 10 mA ion beam 
delivered by the new High
Current Linac at GSI, sophisticated, mainly 
non-intersepting diagnostic devices were developed. Besides the general
set-up  of a versatile test bench, we discuss in particular bunch shape
and emittance measurements. 
A direct time-of-flight technique with a diamond particle detector is used  
to observe the micro-bunch distribution with a  
resolution of $\sim 25 $ ps equals 0.3$^{o}$ in phase.
For the determination of the energy spread a coincidence technique
is applied, using secondary electrons emitted by the ion passing
through an aluminum foil 80 cm upstream of the diamond detector. 
The transverse emittance is measured within one macro pulse with a
pepper-pot system equipped with a high performance CCD camera. 
} 
\end{abstract}

\section{Beam diagnostics for the Linac commissioning}

\begin{figure}[t]
\begin{center}
\includegraphics*[angle=270,width=8cm]{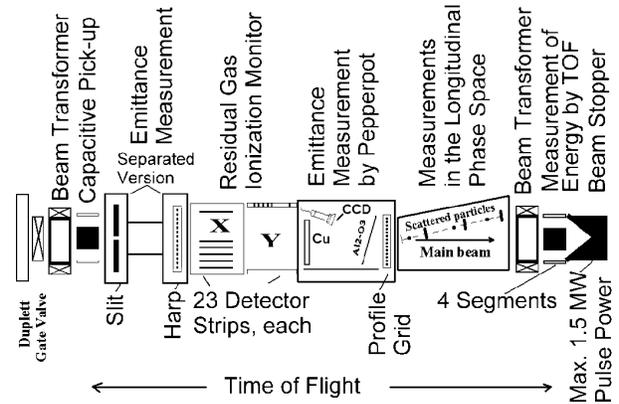}
\caption{Scheme of the test bench as arranged for the 
commissioning of the RFQ. }
\label{BENCH1}%
\vspace*{-0.9cm}
\end{center}
\end{figure}

At GSI upgraded ion sources \cite{reich-linac2000} as well as a new 
36 MHz RFQ- and IH-Linac designed for high current operation 
was commissioned in 1999 \cite{barth-linac2000}. 
New beam diagnostic developments were necessary due to the high beam
power up to  1.3 MW at an energy of 1.4 MeV/u 
within a macro pulse length of maximal 1 ms. 
A beam diagnostics 
bench was installed behind each Linac-structure during the stepwise
commissioning, the scheme is shown in Fig.\ref{BENCH1}. 
Non-destructive devices are used for following tasks:\\
The {\em total current} 
is measured using beam transformer \cite{schneider-biw98}
made of Vitrovac 6065 core
having a $2\times 10$ differential 
secondary winding. The resolution is 100 nA at a bandwidth of 100
kHz. Due to a feedback circuit the droop is less than 1 \% for 5 ms
macro-pulses.\\ 
The {\em beam energy} 
is determined by a  time-of-flight technique using two 50
$\Omega$ matched capacitive pick-ups \cite{strehl-biw00} 
with 1 GHz bandwidth, separated by 2 m. A precision of
$\Delta W /W = 0.1 $ \% is achieved.\\
The {\em beam position} is monitored 
by  digitizing the power of 
the $6^{th}$ harmonics of the rf frequency (216 MHz) of the 4 segments of
these pick-ups.\\
The {\em beam profile} is determined by a residual gas monitor
\cite{strehl-biw00},  where 
residual gas ions are detected on a 23 strip printed board. For typical beam
parameters no significant
broadening of the profiles due to space charge influence is expected.
For lower current or shorter macro pulses conventional profile grids 
are used.

The instruments are now installed behind the last IH2 cavity as well as behind 
the gas stripper.
In the following we discuss the measurement in the longitudinal plane using
particle detectors and of the  transverse emittance using the
pepper-pot system.

\section{Measurement of bunch structure}
\label{section-uboot}

\begin{figure}[t]
\begin{center}
\includegraphics[angle=270,width=5.5cm]{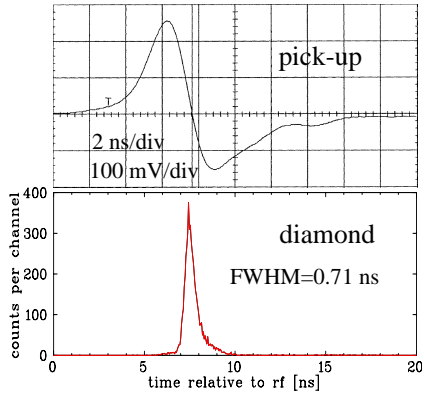}
\end{center}
\vspace*{-0.5cm}
\caption[a]
{Comparison of a pick-up signal to the bunch shape determination using
  the particle detector setup for a 1.4 MeV/u Ar$^{1+}$ beam 3 
m behind the IH2 output. (The 50 $\Omega$ termination of the pick-up 
leads to a  differentiation)}
\label{pick-compare}
\vspace*{-0.1cm}
\end{figure}

\begin{figure}
\begin{center}
\includegraphics[angle=0,width=8cm]{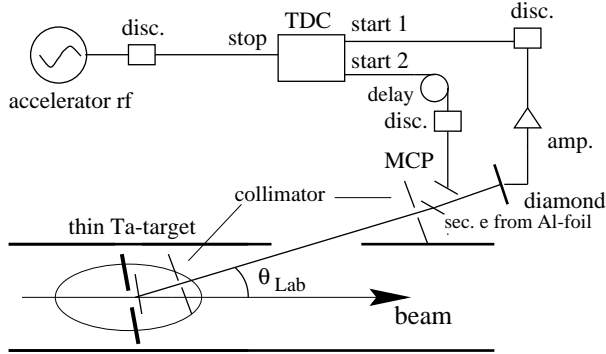}
\end{center}
\vspace*{-0.7cm}
\caption[a]
{Sketch of the designed TOF method 
  for the bunch shape (Sec.\ref{section-uboot}) and phase space
  distribution (Sec.\ref{section-phase-space}) measurement 
with particle detectors.} 
\label{diamant-phasespace}
\vspace*{-0.5cm}
\end{figure}

For the comparison to calculations,
as well as for  matching of different
Linac structures the knowledge of the bunch shape is important, but
measurements are not as common as for the transverse case. At velocities much 
below the speed of light the signal on a transverse
pick-up does not represent the details of the bunch shape due to the large
longitudinal electric field component. A comparison of the pick-up signal to
the bunch shape measured with the method described below is shown in
Fig.\ref{pick-compare} 
for a velocity of $\beta=5.5$ \% (1.4 MeV/u) to
visualize the broadening of the pick-up signal detecting bunches with 
less than 1 ns width.   
We developed 
a device where the arrival of the ion in a particle detector is
measured with respect to the accelerating rf, see
Fig.\ref{diamant-phasespace}.  The method demands for less than one ion hit
per rf period. This reduction is done by Rutherford scattering in a 
210 $\mu$g/cm$^{2}$ tantalum  foil ($\sim 130$ nm thickness) 
at a scattering angle
of $2.5^{o}$ defined by a collimator 
with 16 cm distance and \O   0.5 mm apertures to give a solid angle of 
$\Delta \Omega_{lab}= 2.5\cdot 10^{-4}$. 
The parameters 
are chosen to get an attenuation of $\sim  10^{-8}$ of the beam current. 
A high target mass is preferred, so
the energy spread for a finite solid angle is lower than the
required resolution. For our parameters the largest contribution to the 
energy spread arises from the electronic stopping in the foil, which
amounts e.g. for Ar projectiles to $\sim 0.25$ \%  and
for U to $\sim0.15$ \% (FWHM) at 1.4 MeV/u. 
More details can be found in \cite{forck-dipac99}.

A drawback of this method is the high sensitivity of the tantalum
foil due to the heating by the ions energy loss. Therefore, the beam has to
be attenuated, which  can be done by defocusing. The device is now installed
behind the gas stripper and the first charge separating magnet so that another
type  of attenuation can be applied 
by changing the gas pressure or by selecting a
different charge state. By this means also space charge effects can be
studied. 

Another approach would be the use of a supersonic  high density 
Xenon gas target instead of 
the Tantalum foil; estimations of the effect of the larger elongation have to
be done.

After a drift of $\sim 1$ m the scattered ions are detected by a CVD
diamond detector \cite{forck-dipac99,berder98-bor}.
 Besides the very low radiation 
damage, we gain mainly from the very fast signals, having a rise time below 1
ns. 
The conversion to logical pulses is done by a double
threshold discriminator  \cite{ney98-conf}. 
The logical pulses serve as a start of a VME
time-to-digital converter (CAEN V488), where the stop is derived from the 36
MHz used for accelerating. 
The timing resolution of the system is about 25 ps corresponding
to a phase width of 0.3$^{o}$.

\begin{figure}
\begin{center}
\vspace*{-.5cm}
\includegraphics*[angle=270,width=8.5cm]{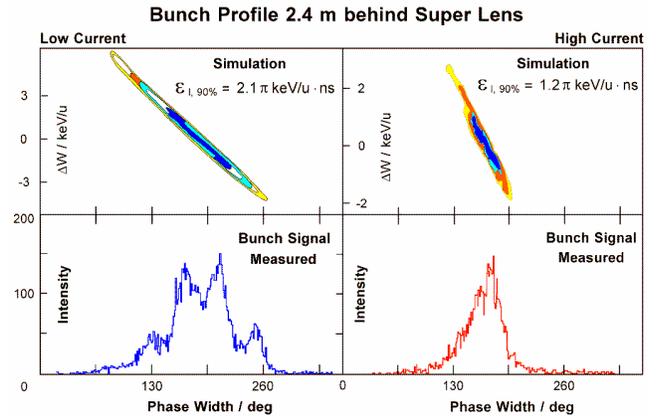}
\end{center}
\vspace*{-0.7cm}
\caption[a]
{Example of the bunch shape measurements 2.4 m behind the
  Super Lens. On the left bottom the result for a low current 0.1 mA Ar$^{1+}$
  beam is shown, on top the calculated emittance is plotted. On the right the
  measurement and simulation are shown for a high current of 5 mA Ar$^{1+}$.} 
\label{sl-variation}
\vspace*{-0.5cm}
\end{figure}

As an example, the bunch structure of a 120 keV/u Ar$^{1+}$ 
beam  at the output of the
Super Lens (and an additional  drift of 2.4 m) 
is shown in Fig.\ref{sl-variation}. The two measurements for low (left) 
and high (right)
current show a quite different bunch 
shape having a larger, filamented emittance
for the low current case.
The particle tracking calculation
\cite{ratz-hif00} shows a strong ion current dependence for the
longitudinal emittance.
The applied rf power in
the cavity counteracts the space charge force for a high current 
beam. For a low current filamentation occurs due to the missing damping by the
space charge.   
Other experimental results can be found in  \cite{barth-linac2000}.

\section{Measurement of longitudinal emittance}
\label{section-phase-space}
The main advantage of using particle detectors is  the possibility to
measure the longitudinal emittance using a coincidence technique. 
As shown in
Fig.\ref{diamant-phasespace}, a second detector can be moved in the path of the
scattered ions. It consists of a 15 $\mu$g/cm$^{2}$ Aluminum foil ($\sim 50 $
nm) where several secondary electrons per ion passage are generated. 
These electrons are accelerated
by an electric field of  1 kV/cm 
towards a micro channel plate equipped with a 50 $\Omega$ 
anode (Hamamatsu F4655-10). 
The time relative to the
accelerating rf is measured as well as the arrival time of the same
particle at the diamond
detector located 80 cm 
behind the MCP. From the difference in time 
of the {\em individual} particles one can generate a  
phase space plot. 

\begin{figure}
\parbox{0.49\linewidth}{
\begin{center}
\includegraphics*[angle=0,width=4cm]{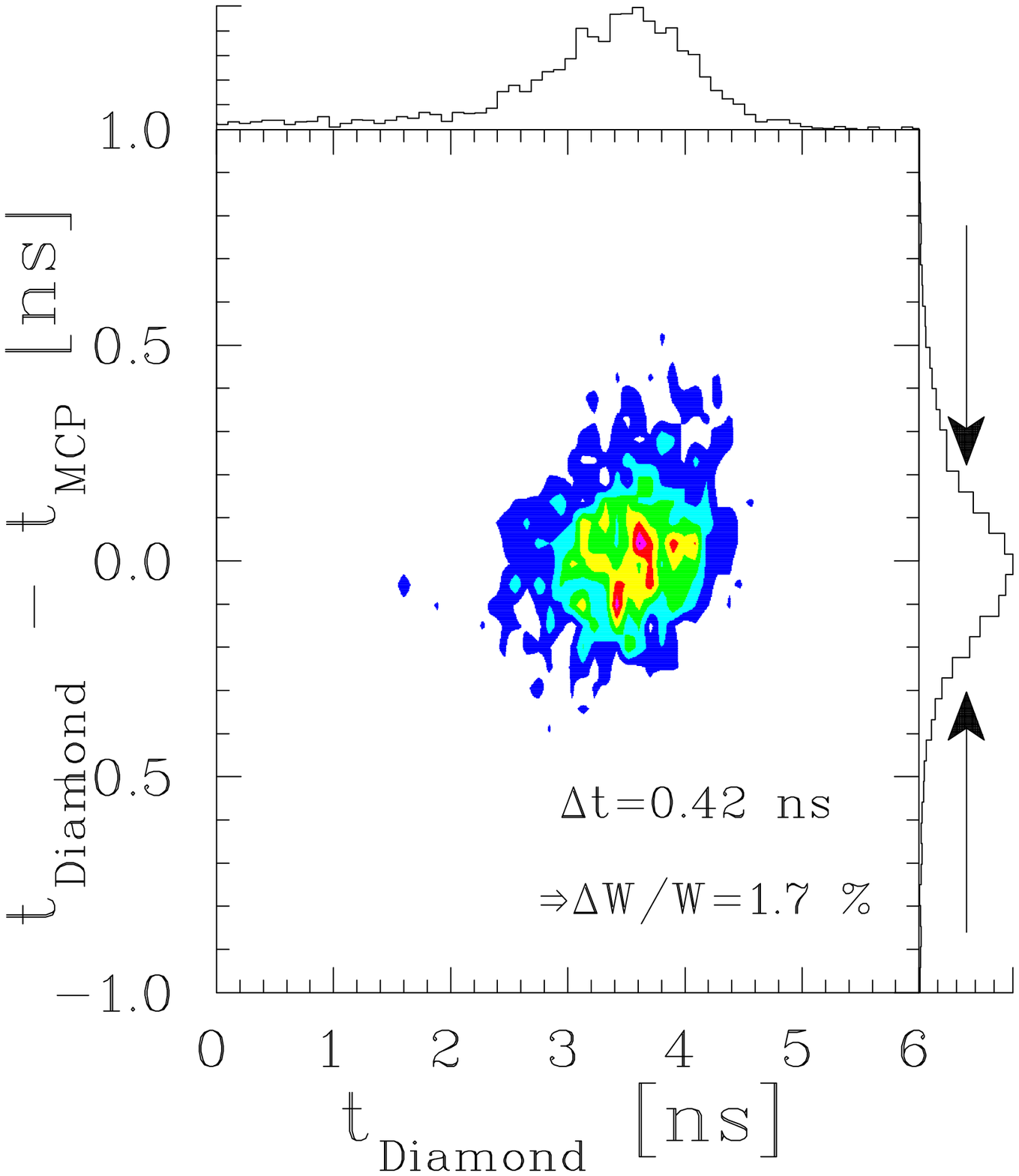}
\end{center}
}
\parbox{0.49\linewidth}{
\begin{center}
\includegraphics*[angle=0,width=4cm]{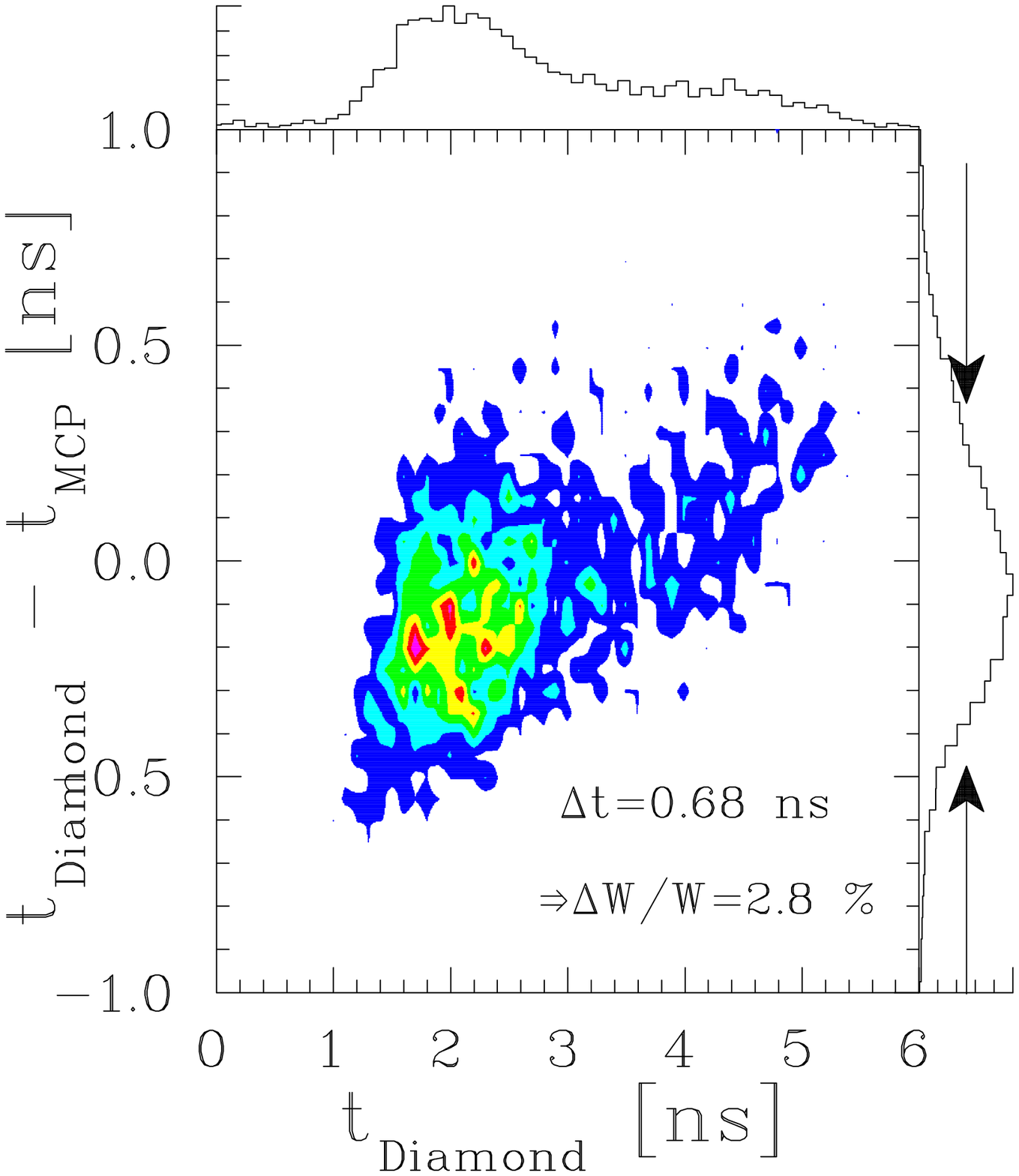}
\end{center}
}
\vspace*{-0.5cm}
\caption[a]
{Measured longitudinal phase space distribution for a 
low current Ar-beam (left) and a high current beam (right) 2.5 m behind the
stripper.  Note that the measured energy spread might 
be too large.} 
\label{coincidence}
\vspace*{-0.5cm}
\end{figure}

An example of such a measurement is given in Fig.\ref{coincidence} (left) 
for a low
current Ar beam 2.5 m 
downstream of the gas stripper.
The arrival times at the diamond detector are used as the
phase (or time) axis having a width of 1.4 ns equals 18$^{o}$ phase width. The 
time difference between diamond and MCP is plotted at the y-axis,  
the width is about 0.4 ns (FWHM)
corresponding to an energy spread of $\Delta W/W= 1.7$ \%. For a high current
beam (5 mA before stripping) a double structure is visible in the bunch
profile 
and an energy broadening to $\Delta W/W= 2.8 $ \% with a 
clear correlation in the phase space. Here the attenuation is done by selecting
a high charge state (Ar$^{15+}$) far from the maximum of the stripping
efficiency curve (Ar$^{10+}$).    
The measured values are larger by  
a factor of 2 as expected by tracking calculation.  Therefore it is
believed that some errors contribute to the measurement: Having a drift
length of only 80 cm between 
the two detectors and an ion  
velocity of 5.5 \% of the ions (corresponding to 48 ns
time of flight)  the accuracy in time
has to be  
25 ps to have a precision of $\Delta W/W$ of 0.1 \%. The imperfections of the
device, in particular the lack of homogeneity of the accelerating field for
the electrons towards the MCP effect the resolution in time. An optimization 
has to be done.  A large distance (e.g. 3 m)
between the two detectors would lower the requirement for the time resolution
of the detectors. Recently it was discovered that there might be some problems 
inside the stripper \cite{barth-linac2000} 
due to inhomogeneity of the gas jet resulting in a wider
energy spread as the design value. 

It is shown, that 
this type of setup can be used for the determination of the longitudinal
emittance at low ion velocities, but a careful design of the components is
necessary.

\section{Measurement of transverse emittance}

\begin{figure}
\begin{center}
\includegraphics*[angle=270,width=5.5cm]{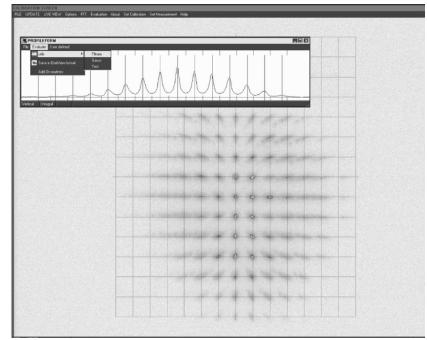}
\end{center}
\vspace*{-0.5cm}
\caption[a]
{Screen shot from the pepper-pot device for an Ar beam and, as an insert, the
  projection onto the horizontal plane.} 
\label{trans-emi}
\vspace*{-0.3cm}
\end{figure}

For the measurement of the transverse emittance two devices were installed at
the diagnostic bench. A conventional slit-grid system \cite{strehl-biw00} 
having a coordinate resolution of 0.05 mm and an angular resolution of 0.3
mrad. Due to the high beam power, this device can only be used for the lower
energy part of the Linac. For high current operation we developed a pepper-pot 
system capable to measure the emittance within one
macro-pulse, see \cite{hoff-biw2000} for more details. 
Here the coordinates are fixed by a $45 \times 45
$ mm$^{2}$ copper plate equipped with $15 \times 15 $ holes with \O 0.1
mm. After a drift of 25 cm the beam-lets 
are stopped on a Al$_{2}$O$_{3}$ screen. The 
divergence of the beam is calculated with respect to the image of the
pepper-pot pattern. This image is created on the screen with  a HeNe laser, 
which illuminate the pepper-pot via a pneumatic driven mirror. This
calibration eliminate systematic errors due to mechanical uncertainties. A
high resolution 12 bit 
CCD camera (PCO SensiCam) transmits the digital data via fiber 
optics. A typical image of such a measurement is shown in Fig.\ref{trans-emi},
together with the projection onto the horizontal or vertical axis. 
This projection is used for the emittance calculation with an algorithm
like for the slit-grid device. 

For a precise measurement the beam width should be large enough to illuminate
several holes in the pepper-pot plate (spacing 3 mm). This also avoids
overheating of the pepper-pot, as well as saturation of the light intensity
emitted from the screen. In addition,  a background level
(about 5 \%) has to be subtracted from the data, probably caused 
by scattered light in the screen. 
Therefore the beam optics have to be changed in some cases to use this modern
and reliable system for a fast measurement in a high current operation.

\begin{figure}
\parbox{0.69\linewidth}{
\begin{center}
\includegraphics*[angle=270,width=4.5cm]{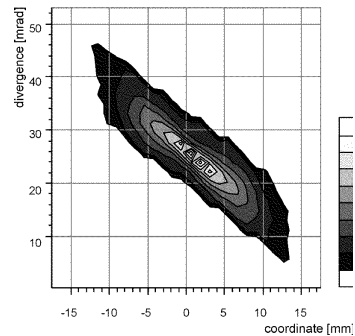}
\end{center}
}
\parbox{0.29\linewidth}{
\caption[a]
{Phase space plot of the data shown above.} 
}
\vspace*{-0.5cm}
\label{trans-emi2}
\vspace*{-0.3cm}
\end{figure}

\end{document}